\documentclass{aa}
\usepackage{psfig}

\def\G{$\Gamma_{\rm x}$ }

\def\approxlt{\mathrel{\hbox{\rlap{\lower.55ex \hbox {$\sim$}}
        \kern-.3em \raise.4ex \hbox{$<$}}}}
\def\approxgt{\mathrel{\hbox{\rlap{\lower.55ex \hbox {$\sim$}}
        \kern-.3em \raise.4ex \hbox{$>$}}}}

\begin{document}

 \title{A two-phase model for the Narrow Line Region of NGC\,4151}


 \author{Stefanie Komossa}       

 \offprints{St. Komossa, \\
  skomossa@xray.mpe.mpg.de}

 \institute{
    Max-Planck-Institut f\"ur extraterrestrische Physik,
         Postfach 1312, D-85741 Garching, Germany  }

 \date{Received:  February 2001; accepted: March 2001}

   \abstract{NGC\,4151 is one of the brightest and best-studied
       Seyfert galaxies. Here, we present a two-phase model of 
       the narrow-line region (NLR) of NGC\,4151. 
       This study is motivated by (i) the fact that the 
       X-ray spectrum of NGC\,4151 is among the flattest known
       for Seyferts, and (ii) the recent {\sl Chandra} 
       detection of an X-ray narrow-line
       region in this galaxy. 
       X-ray spectra as flat as that of NGC\,4151 (\G $\simeq -1.5$)
       are expected to favor the presence  
       of two gas phases in pressure equilibrium (Krolik, McKee \& Tarter 1981).
       In the present study, we show that a
       pronounced two-phase equilibrium develops in the extended emission-line region 
       {\em if we use
       the observed multi-wavelength spectrum of NGC\,4151} to ionize the clouds. 
       The material is stable to isobaric perturbations over
       a wide range of temperatures. 
       We therefore propose that such a condition
       has arisen in the NLR of NGC\,4151, and that it explains
       the detection of hot,
       extended X-ray gas which we identify as the NLR-cloud
       confining medium.  
          } 

        \authorrunning{St. Komossa}
        \titlerunning{A two-phase model of the NLR of NGC\,4151}

   \maketitle

\section{Introduction}

\subsection {NGC\,4151}

NGC\,4151 is the optically brightest Seyfert galaxy 
of type 1.5.
It is very nearby ($r \simeq 20$ Mpc for $H_0$=50 km/s/Mpc)
and is therefore excellently suited 
to study in detail the
physical processes in the nuclei of active galaxies (AGN). 
The extended narrow-emission-line region (NLR) of NGC\,4151 has a cone-like geometry
(Schulz 1988, 1990, Evans et al. 1993, Boksenberg et al. 1995),
interpreted as `radiation cone' (Schulz 1988) caused by anisotropic
illumination of the NLR. In the context of the unified model of AGN  
(Antonucci 1993) such a radiation geometry arises from the   
partial obscuration of a molecular torus or the anisotropic 
emission of the accretion disk.   

NGC\,4151 is the most intensely studied AGN across the electromagnetic
spectrum 
(see Schulz 1995 and Ulrich 2000 for detailed reviews). 
The X-ray spectrum of the galaxy is complex in the soft band (e.g., Weaver et al. 1994)
and extends to very high energies (e.g., Perotti et al. 1981,
Maisack et al. 1993).
The powerlaw spectrum has a mean photon index \G $\simeq -1.5$ 
and is significantly flatter than that of other Seyfert galaxies 
($<$\G$> \simeq -1.9$).  
Using the {\sl Chandra} HETG spectrometer, Ogle et al. (2000)
recently reported the detection of a high-temperature,
narrow-line, X-ray emitting plasma in NGC\,4151, 
confirming earlier {\sl Einstein} (Elvis et al. 1983) and
{\sl ROSAT} (Morse et al. 1995) observations of extended X-ray emission in this
galaxy. The X-ray gas is spatially coincident
with the NLR and extended narrow-line region (ENLR).
A hot intercloud medium was one of several possible explanations
mentioned repeatedly (e.g.,
Elvis et al. 1983, 1990, Morse et al. 1995, Ogle et al. 2000).  
It is therefore important to examine if, and under which conditions,
a two-phase equilibrium of the NLR and ENLR clouds is possible.  

\subsection {Multi-phase cloud models, and motivation
             of the present study} 

Different mechanisms for the confinement of the emission-line  
clouds in AGN have been studied over the last few decades.  The model originally
investigated in detail by Krolik, McKee \& Tarter (1981; KMT hereafter) is a
two-phase model, consisting of cold line-emitting clouds ($T
\simeq 10^4$ K) in pressure balance with a hot inter-cloud medium ($T
\simeq 10^8$ K).  
KMT assumed a relatively flat X-ray spectrum of their input
continuum illuminating the clouds. Later it turned out
that Seyfert galaxies typically show steeper X-ray
spectra and a pressure
balance between a cold, photoionization heated, and a hot, Compton
heated phase then no longer exists (e.g., Fabian et al. 1986).   
More recent studies have demonstrated the presence of an
additional stable region of intermediate temperature (Reynolds \&
Fabian 1995, Komossa \& Fink 1997, Komossa 2001), and 
the recent discovery of warm absorbers located in that
intermediate region has revived the interest in multi-phase cloud
models.  

The aim of this paper, however, is not a general discussion
of the viability of variants of the KMT model to
explain the BLR/NLR-confinement in Seyfert galaxies in general.
Instead, we address the following question:
will a KMT-like two-phase equilibrium develop,
if conditions are favorable in individual galaxies ?  

The observations of extended X-ray emission in NGC\,4151
spatially coincident with the NLR, 
together with the fact that the relatively
hard X-ray spectrum of NGC\,4151 is favorable for the development
of a two-phase equilibrium, motivated the present study.


  \begin{figure}[th]
 \psfig{file=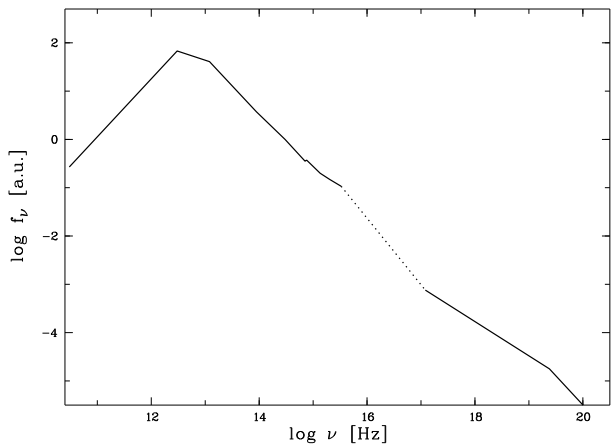,width=8.3cm,clip=}
 \caption[eq]{Average radio to gamma-ray continuum spectrum
               of NGC\,4151, composed from more than 300 
               individual measurements collected from the literature.
            The unobserved EUV part of the spectrum is shown as
              dotted line.  
            The hard X-ray spectrum was conservatively assumed
             to show a break, whereas most measurements indicated an
            extent of the spectrum to much higher energies (see Sect. 3.1.1). 
}
\label{kon}
\end{figure}

\section{Model calculations and results} 

The stability of photoionized gas clouds to isobaric perturbations can be
examined by studying the behavior of cloud temperature $T$ as a function of
pressure.  If the temperature is multi-valued for constant pressure,
and the gradient of the equilibrium curve is positive, several phases
may exist in pressure balance.

The photoionization calculations were carried out with 
the code {\em Cloudy} (Ferland 1993) under the following assumptions:
We adopted solar metal abundances according to Grevesse \& Anders (1989).
The gas density was assumed to be constant ($\log n_{\rm H}=3$),
as was the column density ($\log N_{\rm H} = 21.5$).
Test calculations with different densities and abundances were
performed but they do not alter the results discussed here.  
The clouds were
illuminated by the continuum of a central point-like energy source.
The spectral energy distribution (SED) used for modeling  
corresponds to the {\em average} multi-wavelength continuum of NGC\,4151
(Fig.\,1),
collected from the literature and determined from more than 300
individual data points (Komossa 1994, 
Schulz \& Komossa 1993, and references therein).
The adopted  X-ray spectrum
is particularly important in deriving the shape of the 
equilibrium curve ($T$ versus $U/T$) 
for high temperatures.
We used the mean observed powerlaw index, \G = $-1.5$
(e.g., Perotti et al. 1990), 
extending up to 100 keV, where we assumed the
spectrum to break steeply into the gamma-ray regime
(see Sect. 3.1.1 for further details and references). 
This mean SED is appropriate for modeling,
because the recombination time scale for gas of NLR densities
is long, and the NLR thus sees an average continuum. 

The cloud temperature as a function of the ionization parameter 
was extracted after each model run.
The ionization parameter is defined as 
    \begin {equation}
     U = {Q \over {4\pi{r}^{2}n_{\rm H}c}}~, 
 ~~~{\rm with}~ Q = 4\pi{d}^{2} \int\limits_{\nu_0}^{\infty} 
                     { f_{\nu} \over {h\nu}} \rm{d}\nu~~,
    \end {equation}
where $n_{\rm H}$ is the gas density, $r$ the distance of
the gas cloud from the continuum source, 
$Q$ the number rate of photons
above the Lyman limit $\nu_0$,  
and 
$d$ the distance between observer and galaxy.

Results are shown in Fig.\,2.
We find that a clear two-phase equilibrium between a low-temperature 
and a high-temperature        
component develops over a range of ionization parameters.
We have verified that our results are independent of metal 
abundances{\footnote{For supersolar metal abundances the small region at $T \simeq 10^{6}$\,K
which allows
a {\em three-phase} equilibrium becomes more prominent. However,
photoionization modeling of the ENLR emission-line ratios of NGC\,4151 
(Schulz \& Komossa 1993) 
favored depleted metallicity.}},
gas density, and the amount of absorbed radiofrequency radiation.

In a next step, we added a black-body component to the UV part
of the spectrum. 
A SED with a huge EUV excess
which dominates $Q$ ($Q_{\rm bb} \simeq 0.8 Q_{\rm totl}$) 
was invoked by Penston et al. (1990)
using the ionization-parameter-sensitive emission-line
ratio [OII]/[OIII].    
This SED would significantly narrow down the region
where a two-phase equilibrium is possible (Fig.\,2). 
Since Schulz \& Komossa (1993) showed that the [OII]/[OIII]-ratio
overestimates the ionization parameter (thus $Q$) if the 
emission-line region is inhomogeneous in density,
and since recent studies did not favor a giant EUV excess 
in this galaxy (Alexander et al. 1999), we do not discuss this SED further.  

For comparison, we have plotted in Fig.\,2 the phase diagram obtained
using the average SED of a sample of Seyfert galaxies (Komossa \& Schulz 1997). 
This SED is characterized by a much steeper X-ray
spectrum (\G=$-1.9$) and shows that 
the temperature of the upper branch has reduced and 
the overlap between
upper an lower branch has disappeared.

\section {Discussion}

\subsection{Model assumptions} 

Before we discuss consequences of the models considered here,
we first give some cautious comments on  
the general model assumptions.
The assumptions of photoionization equilibrium and thermal
balance have to be kept in mind, as already discussed by KMT. 
For instance, the gas cooling time scale should be less 
than the dynamical time scale (Morse et al. 1995 
estimate that the time scales are comparable in the ENLR of NGC\,4151).   
The assumption of photoionization equilibrium is problematic
if intrinsic source variability and recombination time scale
are of the same order. 
Given the low mean density of the NLR and ENLR  of NGC\,4151
($\log n \simeq 2-3$) the recombination time scale, 
$t_{\rm rec,H} = \alpha^{-1}(T_{\rm e}) n_{\rm e}^{-1}$,  
is on the order of 1.3\,10$^{2-3}$ yrs, and correspondingly
longer for even lower density. This 
is long compared to the rapid X-ray variability
of this source. An average continuum is thus appropriate for modeling.  
Heating agents in addition to the central continuum source
(like cloud friction) were neglected. KMT showed that their
influence is generally weak. However, radiation pressure by trapped
line photons could significantly increase cloud pressure
(Elitzur \& Ferland 1986).

Apart from these general considerations,
the influence of the {\em shape of the EUV--X-ray continuum specific
to NGC\,4151}
on the results
is particularly important.


  \begin{figure}[t]
 \psfig{file=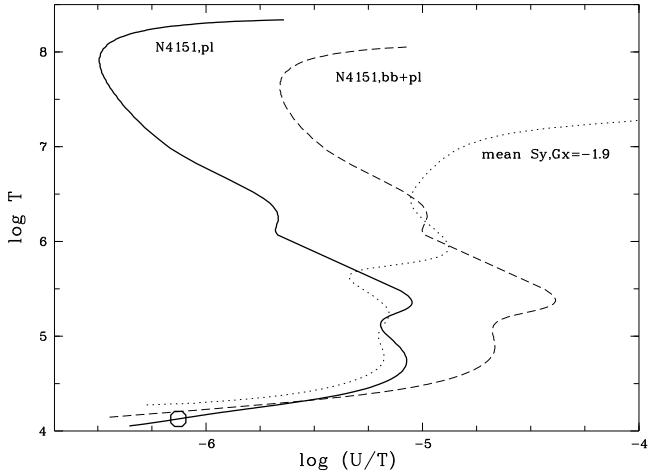,width=8.7cm,clip=}
 \caption[eq]{Equilibrium gas temperature $T$ as a function of $U/T$,
obtained using 
the observed SED of NGC\,4151 (thick solid line).
The open circle marks a value of $\log U = -2.0$. 
Two curves are shown for comparison. 
Dashed line: SED with an additional 
strong EUV-bump component ($Q_{\rm bb}=0.8\,Q_{\rm totl}$; see text);
dotted line: 
mean Seyfert SED
with \G=$-1.9$. 

}
\label{eq}
\end{figure}

\subsubsection {X-ray continuum shape} 

Relevant for the present study is the
shape of the X-ray spectrum (its steepness, and the position
of the break). 
We used an input spectrum with \G = $-1.5$, the mean observed spectral index
(e.g., Perotti et al. 1990){\footnote{for comparison: 
the following photon indices (either means over a collection of
observations, or single measurements) of the hard X-ray
spectrum of NGC\,4151 were reported in the literature:
Beall et al. (1981): \G = $-1.4$, Baity et al. (1984): $-1.6$, 
Warwick et al. (1989): $-1.54$, Maisack and Yaqoob (1991): $-1.37$,
Ogle et al. (2000): $-1.4$; for a collection of \G measurements
of NGC\,4151 prior to 1989 see Tab. 2 of Baity et al. (1984), and 
Tab. 2 of Warwick et al. (1989)}}, 
extending up to 100 keV. At 100 keV the spectrum shows
a break (Maisack et al. 1993) and continues with a significantly steeper
powerlaw. 
It is interesting to note that early X-ray observations of
NGC\,4151 detected the source out to much higher 
energies (e.g., Sch\"onfelder 1978,
Perotti et al. 1981, Perotti et al. 1991),  
whereas some 
recent observations favored a break at 50-100 keV (Apal'kov et al. 1992,
Maisack et al. 1993).  
Were the early detections of the source based on measurement
errors ? We do not believe so, because a few  early
observations {\em did show a break in the spectrum of NGC\,4151}
(Baity et al. 1984, Bassani et al. 1986). In particular, 
Baity et al. presented
several measurements, of which {\em only one} showed a spectral break.
Since taken with the same instrument, but at different epochs,
measurement errors are unlikely, and it is more plausible
that NGC\,4151 underwent real spectral variability.

To be on the `safe side', we conservatively assumed 
the spectrum to break, at $E_{\rm break} = 100$ keV.
Had we adopted a spectrum extending up into the Gamma-regime
(MeV energies) the two-phase equilibrium would have
become even more pronounced, due to the increased
importance of Compton heating.  

\subsubsection {EUV spectrum}

Secondly, the EUV spectral shape is of importance
for the shape of the equilibrium curve.
Using a continuum with a strong EUV excess  
reduces the Compton temperature  of the upper branch 
and thus narrows down the range 
were a two-phase equilibrium is possible
(Fig.\,2; see Fabian et al. 1986, Komossa \& Fink 1997,
Komossa \& Meerschweinchen 2000
for details).

Many different EUV continuum shapes of NGC\,4151, 
and the ENLR emission-line response
to them, were examined by Schulz \& Komossa (1993). 
They concluded that the observed ENLR line-ratios can
be explained by two very different models: 
(i) a SED with a huge EUV bump as suggested earlier
 by Penston et al. (1990) based
on $Q$ estimates, 
and (ii) a powerlaw-like
EUV continuum of much lower $Q$, close to
the lower limit set by the H$\beta$ constraint (and a density-mixed, partly matter-bounded
ENLR, to produce the same observed emission-line ratios).   
A continuum of the latter form (ii) was favored 
by Alexander et al. (1999), 
and adopted by Kraemer et al. (2000),
and we also use it in the present analysis.

\subsection{Two-phase NLR of  NGC\,4151} 

Over the years, a number of different models for the emission-line
regions of AGN have been suggested. Most concentrated on the 
BLR. The broad-line clouds might be confined by
mechanisms other than thermal pressure (e.g. magnetic fields could
play a role; Rees 1987), or they might not be confined at
all. Confinement is unnecessary if the BLR clouds are continuously
produced (e.g., Perry \& Dyson 1985,  
Murray \& Chiang 1995). 
Generally, all suggested models including variants of the
KMT approach, still suffer from some unsolved problems.

In the present study,
it is not our aim to find the ultimate BLR/NLR model
relevant for all types of AGN under any conditions. 
We rather address the question: 
will a KMT-like two-phase equilibrium develop in individual galaxies  
if conditions are favorable ?  
The previous observations of NGC\,4151 are highly suggestive
that this is indeed the case. 
Firstly, the X-ray spectrum of NGC\,4151 is very hard.
Secondly, extended X-ray emission, which could play the role
of the NLR/ENLR confining medium, has already been detected.
The {\sl Chandra} observation revealed similar kinematics of
optical and X-ray NLR which suggests that they are closely linked
(Ogle et al. 2000). 

A number of explanations for the hot extended X-ray emission
of NGC\,4151 have been discussed
(Elvis et al. 1983, 1990, Heckman \& Balick 1983, 
Morse et al. 1995): (i) emission from shocks
due to NLR-cloud -- ISM interaction. However, this would imply a
very small covering fraction of the X-ray gas,
and would not work for the ENLR, since cloud velocities (Schulz 1988) 
are too small.  
(ii) Scattered nuclear emission was considered a possibility, 
but Morse et al. (1995) presented estimates which make this scenario
unlikely. (iii) Thermal emission from a hot confining medium has been 
previously mentioned; and we have shown here that a two-phase medium
develops if illuminated by the observed SED of NGC\,4151. 

Some important consistency checks: 
The mean density of the ENLR, as inferred from the density-sensitive
sulphur emission-line ratio, is $\log n = 2.3$ (Schulz 1988, 
Penston et al. 1990).
There are indications that the density increases inwards 
(e.g., Schulz 1995).  
For the estimates below, we adopt the generally accepted density
scaling law of the NLR, of $n \propto {1\over{r^{2}}}$, with
 $\log n_{\rm cold} = 2.3$ at 1.5 kpc 
(corresponding to $\log n_{\rm hot} = -1.9$; see Fig.\,2)
and $\log n_{\rm cold} = 6.3$ at 15 kpc ($\log n_{\rm hot} = 2.1$).  
What is the column density of the hot inter-cloud medium ? 
The ENLR of NGC\,4151 extends out to $\sim$20\arcsec (=1.9 kpc), 
similar to the distance to which the X-ray emission has been
traced (1.5 kpc, Morse et al. 1995) in the south-west cone. 
We then expect a column density
of $\log N_{\rm H} \simeq 6\,10^{21}$ cm$^{-2}$ of the hot gas.  
The medium is therefore optically thin to Thomson 
scattering  and would not produce any observable effects which 
have not been detected. 
The observed X-ray luminosity of the extended component 
amounts to $\sim$10$^{41}$ erg/s (Elvis et al. 1983: $5\,10^{40}$ erg/s, 
Morse et al. 1995: $3\,10^{41}$ erg/s).  This compares to an X-ray luminosity of
$1\,10^{41}$ erg/s predicted by our model, 
assuming an NLR/ENLR extent and density law as 
given above, and a total filling factor of the gas that corresponds 
to a cone opening angle of 75$^{\rm o}$ (e.g., Evans et al. 1993).  
We note that this comparison should be regarded an order-of-magnitude
estimate, due to uncertainties in both, observations
and modelling assumptions.    
 The hot gas phase is highly ionized and we predict the dominance  
of lines from the ions of, e.g., CVI, NVII, OVIII, SiXIII-XIV, and FeXXIV-XXVI.
Ionization stages up to FeXXV have indeed been detected by {\sl Chandra}.

\section {Summary and conclusions}

The close proximity and brightness of NGC\,4151 
make it a uniquely valuable object to explore
the interaction of the active nucleus with
its immediate environment.  

We have assessed the influence of the observed flat
X-ray continuum of NGC\,4151 on the thermal stability
of NLR/ENLR clouds and the possibility of cloud confinement
by a hot inter-cloud medium.  
We find that a two-phase medium develops and 
we suggest to identify the hot phase with the
observed extended X-ray emission in NGC\,4151,
which is co-spatial with the NLR/ENLR.

We estimate a low density
of the extended gas which is shown to be able
to confine the total NLR and ENLR without becoming
Thomson-thick.    

The results presented here are relevant for other
nearby Seyfert galaxies with relatively hard X-ray
spectra (e.g., NGC\,3227: \G $\approx -1.5$).
Such galaxies are therefore important targets
for future {\sl Chandra} and {\sl XMM-Newton} observations.

\begin{acknowledgements}
We thank Gary Ferland for providing {\em Cloudy},
Hartmut Schulz, Wolfgang Brinkmann and Joachim
Tr\"umper for reading the manuscript,  
and an anonymous
referee for useful suggestions and comments.   
Preprints of this and related papers can be retrieved  
at http://www.xray.mpe.mpg.de/$\sim$skomossa/
\end{acknowledgements}


\begin{thebibliography}{}

\bibitem{} Antonucci R., 1993, ARA\&A 31, 473 

\bibitem{} Alexander T., Sturm E., Lutz D., et al. 1999, ApJ 504, 212  

\bibitem{} Apal'kov Y., Babalyan G., Dekhanov I., et al., 1992, in Proc:
           28th Yamada Conf. on Frontiers of X-ray Astronomy, 28  

\bibitem{} Baity W.A., Mushotzky R., Worral D.M., et al., 1984, ApJ 279, 555

\bibitem{} Bassani L., Butler R.C., DiCocco G., et al., 1986, ApJ 311, 623 

\bibitem{} Beall J.H., Rose W.K., Dennis B.R., et al., 1981, ApJ 247, 458

\bibitem{} Boksenberg A., Catchpole R.M., Macchetto F., et al., 1995, ApJ 440, 151

\bibitem{} Elitzur M., Ferland G.J., 1986, ApJ 305, 35 

\bibitem{} Elvis M., Briel U., Henry J.P., 1983, ApJ 268, 105   

\bibitem{} Elvis M., Fassnacht C., Wilson A.S., Briel U., 1990, ApJ 361, 459 

\bibitem{} Evans I.N., Tsetanov Z., Kriss G.A., et al., 1993, ApJ 417, 82 

\bibitem{} Fabian A.C., Guilbert P.W., Arnaud K.A., et al.,  
                      1986, MNRAS 218, 457

\bibitem{} Ferland G.J., 1993, University of Kentucky, Physics Department, 
Internal Report 

\bibitem{} Grevesse N., Anders E., 1989, in `Cosmic Abundances of Matter',
                        C.J. Waddington (ed.), AIP 183, 1 
                        (New York: American Institute of Physics) 

\bibitem{} Heckman T.M., Balick B., 1983, ApJ 268, 102 

\bibitem{}Komossa S., 1994, Diplom Thesis, Ruhr-Universi\-t\"at Bo\-chum

\bibitem{}Komossa S., 2001, in Proc: IX. Marcel Grossmann Meeting 
 on General Relativity, Gravitation and Relativistic Field Theories,
 V. Gurzadyan et al. (eds), in press [astro-ph/0101289]

\bibitem{}Komossa S., Fink H., 1997, A\&A 322, 719 

\bibitem{}Komossa S., Meerschweinchen J., 2000, A\&A 354, 411

\bibitem{}Komossa S., Schulz H., 1997, A\&A 323, 31 

\bibitem{}Kraemer S.B., Crenshaw D.M., Hutchings J., et al., 2000, ApJ 531, 278 

\bibitem{} Krolik J.H., McKee C.F., Tarter C.B., 1981, ApJ 249, 422  

\bibitem{} Maisack M., Johnson W.N., Kinzer R.L., 1993, ApJ 407, L61 

\bibitem{} Maisack M., Yaqoob T., 1991, A\&A 249, 25 

\bibitem{} Morse J., Wilson A.S., Elvis M., Weaver K., 1995, ApJ 439, 121 

\bibitem{} Murray N., Chiang J., 1995, ApJ 454, L105           

\bibitem{} Ogle P.M., Marshall H.L., Lee J.C., 
             Canizares C.R., 2000, ApJL in press [astro-ph/0010314] 

\bibitem{} Penston M.V., Robinson A., Alloin D., et al., 1990, A\&A 236, 53  

\bibitem{} Perotti F., DellaVentura A., Villa G., et al., 1981, ApJ 247, L63

\bibitem{} Perotti F., Buratti R., Maggioli P., et al., 1990, ApJ 356, 467 

\bibitem{} Perotti F.,  Maggioli P., Quadrini E., et al., 1991, ApJ 373, 75 

\bibitem{} Perry J.J., Dyson J.E., 1985, MNRAS 213, 665 

\bibitem{} Rees M.J., 1987, MNRAS 228, 47 

\bibitem{} Reynolds C.S., Fabian, A.C., 1995, MNRAS 273, 167  

\bibitem{} Sch\"onfelder V., 1978, Nature 274, 344 

\bibitem{} Schulz H., 1988, A\&A 203, 233 

\bibitem{} Schulz H., 1990, AJ 99, 1442 

\bibitem{} Schulz H., 1995, Habilitationsschrift, Ruhr-Universi\-t\"at Bo\-chum  

\bibitem{} Schulz H., Komossa S., 1993, A\&A 278, 29  

\bibitem{} Ulrich M.-H., 2000, A\&ARv 10, 134 

\bibitem{} Warwick R.S., Yaqoob T., Pounds K.A., 1989, PASJ 41, 721 

\bibitem{} Weaver K., Mushotzky R.F., Arnaud K.A., et al., 1994, ApJ 423, 621 

\end{thebibliography}
\end{document}